%% file: a_MAIN.tex
\documentclass{article}

\usepackage{arxiv}

\usepackage[utf8]{inputenc} 
\usepackage[T1]{fontenc}    
\usepackage{hyperref}       
\usepackage{url}            
\usepackage{booktabs}       
\usepackage{amsfonts}       
\usepackage{nicefrac}       
\usepackage{microtype}      
\usepackage{lipsum}
\usepackage{graphicx}
\usepackage{multirow}
\usepackage{hanging}

\usepackage[table]{xcolor}
\definecolor{darkred}{rgb}{0.8, 0.0, 0.0}

\definecolor{nicegreen}{RGB}{76, 175, 80}

\title{Can Telematics Improve Driving Style?\\
The Use of Behavioural Data in Motor Insurance
}

\author{
 Alberto Cevolini\\
  Dept of Political and Social Sciences\\
  University of Bologna\\
\And
 Elena Morotti \\
  Dept of Political and Social Sciences\\
  University of Bologna\\
\And
 Elena Esposito \\
 Dept of Political and Social Sciences, University of Bologna\\
 Faculty of Sociology, Bielefeld University and\\ 
\And
 Lorenzo Romanelli \\ Swiss Re Europe S.A., Rappresentanza per l'Italia
\And
 Riccardo Tisseur \\ Swiss Re Europe S.A., Rappresentanza per l'Italia
\And
 Cristiano Misani \\ Swiss Re Europe S.A., Rappresentanza per l'Italia
}

\begin{document}
\maketitle

\begin{abstract}
Motor insurance can use telematics data not only to understand the individual driving style, but also to implement innovative coaching strategies that feed back to the drivers, through an app, the aggregated information extracted from the data. The purpose is to encourage an improvement in their driving style. Precondition for this improvement is that drivers are digitally engaged, that is, they interact with the app. Our hypothesis is that the effectiveness of current experimentations depends on the integration of two distinct types of behavioural data: behavioural data on driving style and behavioural data on users’ interaction with the app. Based on the empirical investigation of the dataset of a company selling a telematics motor insurance policy, our research shows that there is a correlation between engagement with the app and improvement of driving style, but the analysis must distinguish different groups of users with different driving abilities, and take into account time differences. Our findings contribute to clarify the methodological challenges that must be addressed when exploring engagement and coaching effectiveness in proactive insurance policies. We conclude by discussing the possibility and difficulties of tracking and using second-order behavioural data related to policyholder engagement with the app.
\end{abstract}


\input{aa_Introduction}

\input{aa_EvolutionOfUbi}

\input{aa_PreviousResearch}

\input{aa_DatasetMethodology}
\input{aa_Results}

\input{aa_Concl}



\bibliographystyle{plain}
\bibliography{biblio}  

\end{document}

%% file: aa_Introduction.tex
\section{Introduction}\label{sec:intro}

Insurance is a fundamental risk-transfer mechanism of modern society. The risks of the insured are financially \textit{transferred} to the insurer and at the same time \textit{transformed}: the damage that would be financially ruinous for an individual is distributed by the insurer among all the members of the same pool and thus becomes sustainable \cite{walters1981risk, albrecht1990risikotransformationstheorie, farny2011versicherungsbetriebslehre}. Despite the effectiveness of this risk pooling and spreading mechanism, insurers have an interest in preemptively reducing the risks transferred by the insured. Since many risks depend on individual behaviour, as in the illustrative case of motor insurance, the insurer’s preventive activity should move from, and act upon, the behaviour of the insured. But policyholders’ behaviour is in principle unobservable.

Since the early 2000s, insurance companies selling third-party liability motor insurance policies have invested heavily in the use of telematics data to track drivers’ behaviour. The data collected should make it possible to assess policyholders’ risk profile and adjust their policy premium accordingly. The insurance industry terms this opportunity cusage-based-insurance' (UBI).

Business experience over the last two decades shows a significant evolution in the use of behavioural data. Insurance companies not only use it to refine the risk profile of policyholders but also feed the aggregated information they obtain from behavioural data back to the drivers. The aim is to promote greater awareness by policyholders of their driving style and to encourage a change in driving behaviour in the case their driving habits show criticalities that increase risk exposure. In this respect, insurance companies speak of \textit{coaching} \cite{Donaldson2020, tanninen2021tracking,cevolini2022coaching}. 
The usual abstract assumption is that coaching works because policyholders who receive feedback improve their driving style and become better drivers \cite[p.~22]{soleymanian2019sensor}.

The effectiveness of coaching, however, is yet to be proven and depends on a crucial factor: the propensity of policyholders to take in and use the information fed back by the insurance company in order to motivate them to change their behaviour. This depends, in turn, on the willingness of users to consult the tools that convey this information, which usually are digital apps. Based on a rapidly developing strand of research \cite{lupton2013digitally, asimakopoulos2017motivation, Bucher2020engaged}, we suggest calling this kind of activity \textit{engagement}.
Engagement can be understood in either a broad or narrow sense. A recent paper \cite{Discovery2022} uses the term in a broad sense: policyholders who took out health and life insurance based on behavioural data are considered ‘engaged’ when they participate in pre-established programs including diagnostic screening, gym membership, and tracked daily exercise to promote a healthy lifestyle.

In our research on telematics motor insurance, instead, we use the term ‘engagement’ in a narrow sense to refer exclusively to the users’ interaction with the app. We focus therefore on the interplay of information feedback and behavioural change. Our research assumes that engagement is also a behaviour, which like driving behaviour is (or can be) tracked by digital devices. If by ‘behaviour’ we do not only mean ‘the use of the car’ but also ‘the use of the app’, then insurance companies have to deal with two different types of behavioural data—behavioural data on driving style, and behavioural data on users’ interaction with the app.

According to this approach, the usual UBI formulation should be clarified. The success of proactive strategies implemented by insurance companies is, in fact, based on two different types of behaviour that can both be tracked by the telematics app: not only the driving behaviour of policyholders but also their interaction with the app, i.e., their being engaged. 
Insurance companies selling telematics insurance policies collect a lot of data about both behaviours: use of the car and use of the app. This is, in our opinion, a crucial novelty.
Behavioural data are considered a ``remarkable advance'' in automobile insurance \cite[p.~662]{guillen2019use}. 

Previously, the insurance industry could only use variables related to fixed characteristics of the policyholder and the vehicle, many of which, such as age and gender, are not causally related to the risk of getting into a crash. They are proxy variables.
Behavioural variables, instead, are \textit{causally} related to the risk of road accidents and promise to enable personalised tarification, which may be considered a fairer policy premium setting system \cite{meyers2018enacting, cevoliniesposito2022from}. Moreover, as we have seen, behavioural data processing can be carried out to implement coaching strategies and possibly improve policyholders’ driving behaviour. Behavioural data processing, thus, is expected to impact on policyholders (who know that their behaviour is monitored), on insurance companies (which can improve their predicting capacity by combining behavioural and non-behavioural variables), and on the relationship between policyholders and insurance companies (triggering feedback loops).

Our hypothesis is that the effectiveness of current experiments depends on the integration of these two distinct types of behavioural data. This integration raises a number of new questions: How should engagement be properly defined? How should it be measured? Is there any empirical evidence of a connection between engagement and driving behaviour improvement? And how does this connection change over time?
To answer these questions, we investigated the dataset of an insurance company selling telematics motor insurance policies.
In the next section, we describe the emergence of the idea of insurance as a loss prevention institution and the evolution of usage-based auto insurance policies over the past two decades. In section \ref{sec:PrevResearch}, we provide a brief overview of relevant research. In section \ref{sec:DatasetMethod}, we describe the dataset we worked on, the methodology we followed, and some limitations of our study. 
Section \ref{sec:results} presents our main findings. Section \ref{sec:Concl} exposes our conclusion and suggests possible directions for future developments of behavioural insurance.

%% file: aa_EvolutionOfUbi.tex
\section{The evolution of usage-based auto insurance policies}\label{sec:EvolutOfUBI}

In the mid-1990s, the motor insurance industry began to question the insurance model that merely compensates policyholders’ claims. Starting from the assumption that the majority of road accidents are caused by human miscalculations (of driving capability, road conditions, or driving control under certain road and weather conditions), the possibility of insurance companies acting as loss prevention companies began to be discussed. The aim was ``topping claims before they happen'' \cite[p.~271]{ericson2003insurance}.
Underpinning this project was the conviction that insurance could not simply be a risk spreading mechanism. Spreading risks basically means that policyholders transfer their risks to the insurance company, which distributes them over the pool of insured customers. The result is a risk mitigation for the customers who feel relieved from the financial consequences of possible future damages. A consequence, however, can also be that policyholders are less incentivized to take precautionary measures, producing the thorny problem of moral hazard \cite{heimer1985reactive, stone2002beyond}.
To counter this attitude, it has been suggested to try ``to make people more individually accountable for risks'' \cite[p.~1]{baker2002embracing}. The basic idea was to move from ``spreading risks'' to ``embracing risks'': even if policyholders pay for coverage, they should be aware that they retain, at least in part, both a moral and financial responsibility for the consequences of their behaviour \cite[p.~3]{baker2002embracing}.
In the case of auto insurance, this meant that drivers should engage in preventive actions. But prevention first requires an awareness of the risks to be avoided in order for bad driving habits to be removed \cite[p.~278]{ericson2003insurance}. What remained unclear, however, was how the insurance industry could tackle the problem of bad driving. This is where digital devices used as monitoring devices to produce behavioural data come into play.

The first form of \textit{usage-based-insurance} (UBI) tested in the early 2000s was the so-called \textit{pay-as-you-drive} (PAYD) insurance policy \cite{litman2005pay}. The novelty of this policy was that its pricing system was based on the mileage driven by the policyholders during the policy term. The underlying idea was that mileage is a crucial risk factor statistically related to claim probability. The assumption was that people with low mileage are low-risk motorists and should pay less, whereas people with high mileage are high-risk motorists and should pay more.
The PAYD-pricing system was later questioned as it does not take into account that higher mileage can also mean higher driving experience producing better driving skills. Increasing mileage can be connected with a 'learning effect' \cite{guillen2019use} that, in turn, might decrease the risk of road accidents. Between young licensed drivers and claim probability, on the other hand, there is a similar statistically significant relationship.\\
UBI has later evolved into \textit{pay-how-you-drive} (PHYD) insurance policies, based on the idea that driving style is causally related to the risk of road accidents and should also be taken into consideration when setting the policy premium. Between statistical variables like gender and age and claim probability there is actually a strong statistical correlation but no evident causal relationship. Between phone distraction and the likelihood of getting into a crash, instead, there is a causal relationship. PHYD insurance policies, therefore, keep measuring mileage as PAYD policies, but they also track drivers’ behavioural characteristics to assess their actual driving style.
Drivers’ behaviour is tracked by means of telemetry packages. PHYD insurance policies usually require the installation of a black box in the car with policyholder’s consent. This black box generates a huge amount of behavioural data that allows the company to monitor the policyholders’ driving style—how they steer, how and how often they brake, whether they exceed the speed limit, whether they drive predominantly during the day or at night, and so on. The aggregation of these features makes it possible to assess the individual risk profile and can be used to adjust the policy premium accordingly. This information can also be the basis for coaching services that aim to prevent claims before they occur \cite{guillen2021using}.

The crucial condition to implement coaching strategies is \textit{feedback}. In most advanced telematics insurance solutions, drivers who take out a PHYD insurance policy are supposed to download an app on their smartphone. This app notifies policyholders of the overall score they achieved depending on how good or bad they drove. The same app also communicates the scores achieved in the main features (manoeuvres) that the company uses to define the individual driving profile \cite{Discovery2020, romanelli2021driver}.  Finally, the app shows every single trip travelled by the insured and indicates exactly whether any criticalities have been found and what they are (e.g., where the insured exceeded the speed limit or made a U-turn). This information is made available after driving, not real-time.\\
By means of feedback, information literally \textit{circulates}, that is, it runs circularly. Drivers disclose information about their driving behaviour to the insurance company. The insurance company, in turn, discloses information concerning risk assessment and risk profile to the drivers. Telematics insurance policies, thus, do not simply turn information asymmetry upside down, as many scholars argue \cite{siegelman2014information, lasry2015rencontre}. They rather trigger a circular relationship where behaviour produces information, and information is fed back to change behaviour. What is really going on in PHYD insurance policies is a kind of `feedback loop'.

%% file: aa_PreviousResearch.tex
\section{Previous research}\label{sec:PrevResearch}

As shown by a recent bibliometric review of telematics-based auto insurance \cite{chauhan2024bibliometric}, the literature on telematics motor insurance is very large and ever-expanding. Here, we only focus on the contributions which explore the relationship between information feedback and driving behaviour. A recent overview of studies investigating the impact of telematics on road safety points out that there is still scarce research about before/after feedback provision to the drivers \cite{ziakopoulos2022transformation}.

More than twenty years ago, Wouters and Bos \cite[p.~644ff]{wouters2000traffic} put forward the hypothesis that drivers who know that they are being monitored might be encouraged to change their behaviour, especially if they receive feedback as a result of this monitoring. In their empirical research on a business fleet, Wouters and Bos assessed the effect of this ‘behavioural feedback’ based on JDR (journey data recorder) by comparing an experimental group with a control group of vehicles for a period of 12 months. A statistically significant accident reduction could be detected only for some of the fleet sets, but the overall accident rate in the experimental group was reduced by 20\% after the intervention. However, both monitoring and behavioural feedback were not linked to an insurance policy and lacked the reinforcement that insurance policies usually provide in addition to feedback, namely, financial incentives. 

A decade later, Farmer, Kirley and McCartt in \cite{farmer2010effects} tested the effects of in-vehicle monitoring on the driving behaviour of teenagers, whose crash rates, as well-known, is consistently higher than any other age group. Feedback, in this case, was notified to their parents on a dedicated website. After 24 weeks monitoring on 85 recently licensed drivers in a suburban Washington DC area, it turned out that there were no statistically relevant changes in driving behaviour and that parents themselves made few visits to the website to check the driving behaviour of their children. Also in this case, feedback was not associated to an insurance policy. 

In 2011, Bolderdijk, Knockaert, Steg and Verhoef in \cite{bolderdijk2011effects} carried out a field experiment on the effects of a PAYD insurance policy on young drivers’ speeding behaviour. The basic reasoning was that young drivers are overrepresented in road accidents statistics because they tend to drive at higher speed, and speed is one of the most important behavioural determinants of crash risk. The goal of their research was to test if the provision of financial rewards for keeping the speed limit could encourage young drivers to modify their driving behaviour. Participants could check their performance by logging in to a website which provided detailed feedback on speed violations, mileage and night-time driving, and showed by default the prospective overall discount they could earn. The incentive group (ca. 150 participants) showed a modest but significant reduction in speeding, strongly associated to financial incentives (when financial incentives were removed, speeding increased again).

The research that comes closest to the problems we investigate in this article is that of Soleymanian, Weinberg and Zhu \cite{soleymanian2019sensor}. Based on the dataset provided by a major US insurance company offering a PHYD policy, Soleymanian and colleagues were able to observe more than 100,000 customers over a 32-months period. Their main research question was whether there is a statistically significant improvement in driving behaviour of UBI customers compared to usual customers. Their research showed that UBI customers improved their driving score by ca. 9\%  (from 62.05 in week 1 to 67.87 in week 26), that this improvement was higher in early weeks and for young drivers, and that it did not depend solely on feedback but also on financial incentives. 15\% of UBI customers dropped out in weeks 11 and 12. The consequence was a significant decline in harsh braking, that can be interpreted as the outcome of self-selection: PHYD policies retain best customers and let bad customers leave. 

Soleymanian, Weinberg and Zhu’s research \cite{soleymanian2019sensor} is very important, mainly because it is based on an insurance company dataset and observes UBI customers over time. However, it does not investigate many issues that are crucial for us. For example, if the purpose of coaching strategies is the improvement of policyholders’ driving behaviour, how should an improvement be properly defined? And how should it be measured by insurance companies that have access to behavioural data? Since improvement should be the result of coaching, how should a coaching process be defined? Are there short-term and long-term coaching effects? Crucial questions for us are also how engagement can be defined and measured, whether there is a connection between engagement and driving behaviour improvement, and how this connection changes over time. In our empirical research, we deal with these questions.

%% file: aa_DatasetMethodology.tex
\section{Dataset and methodology}\label{sec:DatasetMethod}

The data we worked on were taken from PHYD insurance policies based on mobile telematics. In this case, an app in the smartphone replaces the usual black box. Such a replacement has advantages and disadvantages. The smartphone is usually regarded as an excellent platform for providing users with prompt feedback \cite{handel2014insurance, wahlstrom2015driving}. Moreover, the smartphone detects phone distraction, which is known to be one of the main causes of road accidents and a crucial feature to be integrated into the score evaluation process. The main disadvantage is that telematics data produced by a smartphone are less accurate than the telematics data produced by a black box and require more pre-processing.

During each trip, the app we analysed downloads geolocation information from its map provider and records raw data from the GPS, the accelerometer, and the gyroscope as well as from the smartphone system to check for phone usage. The data is processed in real-time to assess each driving session based on four key aspects: the attention paid by the driver who should not use the phone while driving the car, compliance with speed limits, cautiousness on the road, and a set of circumstances depending on some external factors (such as driving time during rush hours). Accordingly, four sub-scores are generated to analyse each trip feature separately, namely: ‘attentive driving’, ‘conscious driving’, ‘smooth driving’, and ‘contextual’ scores. They all range from 0 (poor) to 100 (excellent) and are described in Table \ref{tab:ColorideFeat}.

\begin{table}[ht]
    \centering
    \begin{tabular}{ p{2cm} p{8cm} p{5cm}}
    \toprule
    Sub-score     & Description & Inputs \\
    \midrule
Attentive driving     
    &   It evaluates the driver’s level of attentiveness by assessing distractions caused by mobile phone usage. A higher frequency of phone usage events per kilometer results in a lower score for attentive driving.
    & Count of phone unlock events; total driven kilometres. \\
    \arrayrulecolor{black!30}\midrule
Conscious driving   
    &   It evaluates the driver’s adherence to speed limits by assessing both the frequency and magnitude of speed limit violations. A higher percentage of the trip spent driving above speed limits, as well as larger differences between posted speed limits and actual driving speeds when exceeding limits, result in a lower score for conscious driving.
    & Total driven time; driven time above speed limits; difference between driving speed and speed limit. \\
    \arrayrulecolor{black!30}\midrule
Smooth driving
    & It evaluates the driver’s level of cautiousness by analyzing the frequency of risky maneuvers during the trip. Risky maneuvers include harsh braking, acceleration, cornering, steering, failure to yield at intersections, aggressive roundabout maneuvers, and U-turns. A higher frequency of risky maneuvers per kilometer results in a lower score for smooth driving.
    & Count of detected risky manoeuvres; total driven kilometres. \\
    \arrayrulecolor{black!30}\midrule
Contextual 
    & It evaluates the potential impact of external factors on driving behaviour, by considering the amount of time spent driving in risky contexts. Risky contexts may include driving during rush hours, at night, or on urban roads. A greater amount of time spent driving in these risky contexts results in a lower score for the contextual sub-score.
    & Total driven time; driven time during night hours; driven time during rush hours; driven time on urban roads. \\
     \arrayrulecolor{black}\bottomrule
    \end{tabular}
    \caption{Description of the four sub-scores assessing each driving session.}
    \label{tab:ColorideFeat}
\end{table}

\begin{figure}
    \centering
    \includegraphics[width=0.75\linewidth]{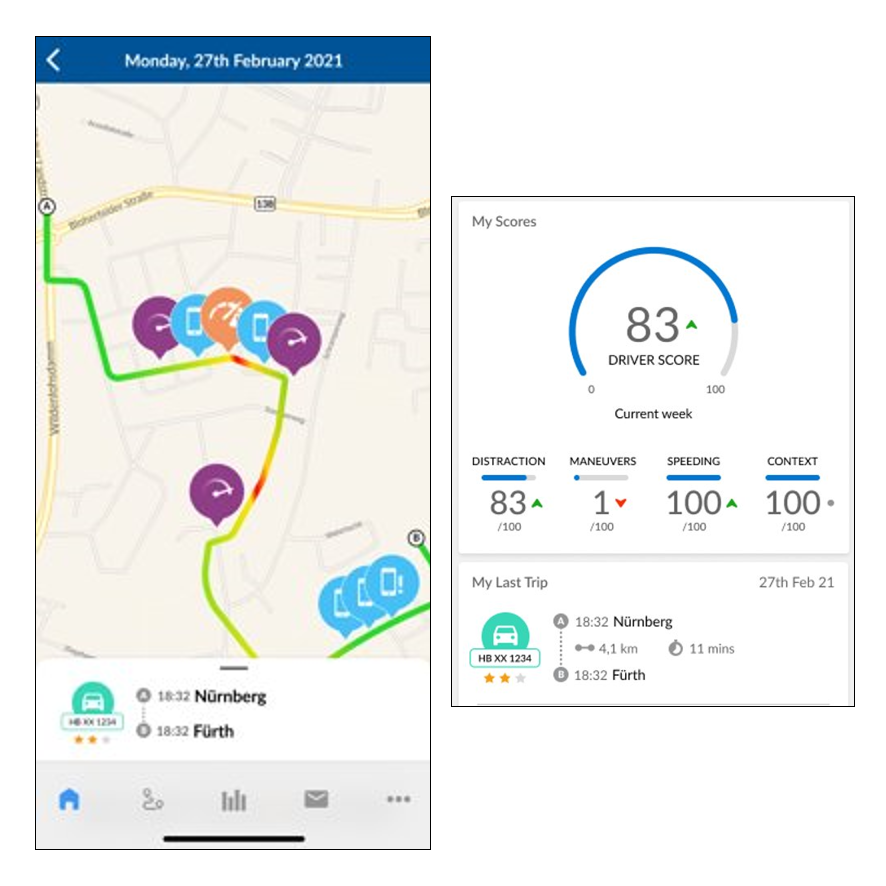}
    \caption{On the left, the critical events displayed on each trip map, such as phone distractions (blue icons), risky manoeuvres (purple and orange icons), and speeding (highlighted as red road segments). On the right, the app homepage where the weekly overall driver score and the four sub-scores are updated after each trip.}
    \label{fig:Coloride}
\end{figure}

A cumulative trip score is computed as a weighted average of the four sub-scores. It still ranges between 0 and 100, but weights are set by the insurance company, which can balance the importance of the features as it considers most appropriate. Immediately after each driving session, the trip scores are displayed, and all significant critical events are localized on the trip map (as shown by the first screenshot in Figure \ref{fig:Coloride}) to ease the scores interpretation. Past trips remain visible and searchable in the app, while a weekly score for the current week is updated on the app homepage, as visible on the right image of Figure \ref{fig:Coloride}.

In many mobile telematics-based insurance policies on the market, the score is used to reward policyholders with financial incentives (e.g., fuel cashback, vouchers, discount upon renewal of the policy). This is not the case of the UBI motor insurance policy we worked on. On the one side, the absence of financial rewards gave us the opportunity to investigate the pure interplay of information feedback and behavioural change. On the other side, the absence of financial rewards deprived us of the possibility of exploring the relationship between digital engagement and a bonus system based on behavioural improvement. As we explain in the \ref{subsec:limitations}, this is an important limitation of our research.

\subsection{ Data Preprocessing}\label{subsec:DataPreproc}

\begin{figure}
    \centering
    \includegraphics[width=0.9\linewidth]{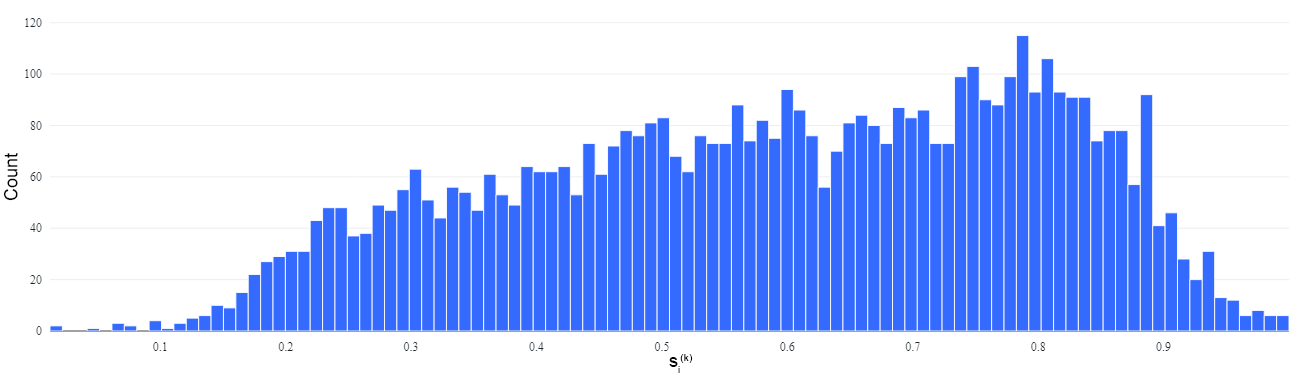}
    \caption{Histogram of all weekly scores.}
    \label{fig:hist_WeeklyScore}
\end{figure}

We accessed trip scores and trip metadata of 498 new customers, onboarded in a 9-month period from March 2022 in a western European country. The observation period was 35 weeks. 
In the app weekly summary, the automatically defined week always starts on Monday, disregarding the exact onboarding day of each individual. For this reason, we considered Monday-starting weeks and associated the week indexes $i = 0, 1, ...$ to each trip, where $i = 0$ corresponds to the onboarding week for each user, 'aligning' the users according to their own timing. 
Coherently, for each user $k \in \{1, ..., 498\}$, we have computed the weekly score $s_i^{(k)}$ at each $i$-th week as the mean of the trip scores of that week. All the values have then been linearly scaled from the $[0, 100]$ range into $[0, 1]$,  hence each $s_i^{(k)}$ takes values from $0$ (bad) to $1$ (excellent). The total number of weekly scores is 4419. This feature is described in the histogram in Figure \ref{fig:hist_WeeklyScore} and in the first row of Table \ref{tab:EdaWeeklyData}. The median score is 0.6065 and half of the values are between 0.4343 and 0.7690.

\begin{table}
 \caption{Descriptive statistics of aggregated weekly data.}
  \centering
  \begin{tabular}{l rrrrr} 
    \toprule
    Feature                         & Minimum & Q1 & Q2 & Q3 & Maximum             \\
    \midrule
    Weekly Score                    & 0.0120 & 0.4343 & 0.6065 & 0.7690 & 1      \\
    Weekly app session duration (secs)          & 0 & 0  & 12  & 186 & 4813  \\
    Number of weekly app sessions   & 0 & 0 & 1 & 3 & 54 \\
    \bottomrule
  \end{tabular}
  \label{tab:EdaWeeklyData}
\end{table}

\begin{figure}
    \centering
    \includegraphics[width=0.9\textwidth]{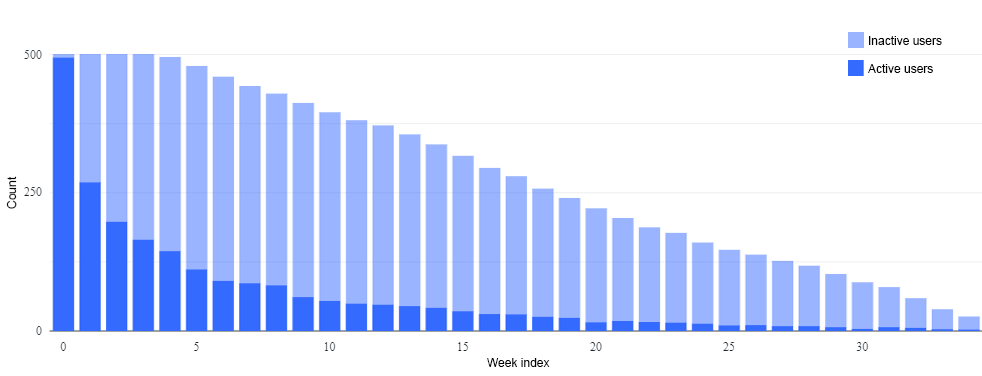}\\
    \includegraphics[width=0.9\textwidth]{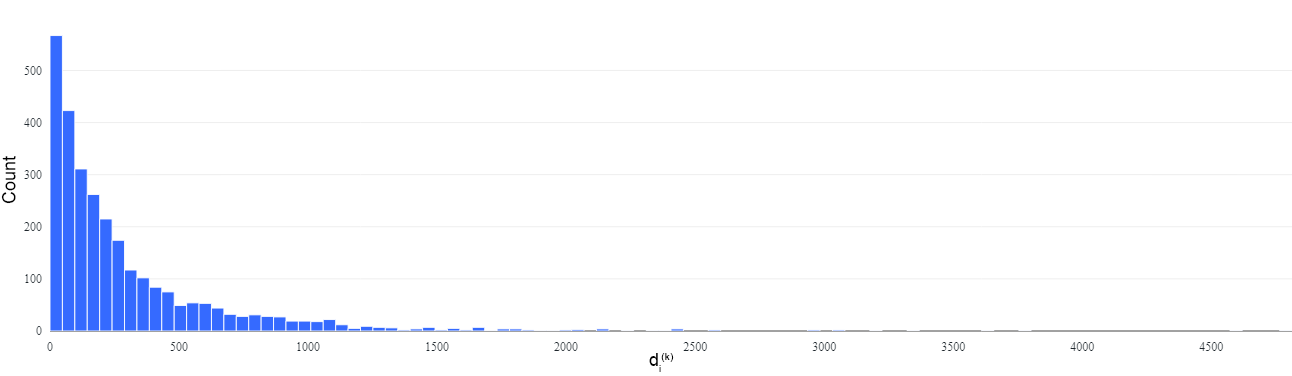}
    \caption{On the top, visualization of the number of active users on the telematics application (i.e., with at least one app session during the week) and inactive users (with no app sessions), as a function of the week index. On the bottom, histogram of the computed weekly durations (in seconds).}
    \label{fig:EDA_engagem}
\end{figure}

Telematics devices, however, can also collect further behavioural features that do not (yet) go into the score. Our app also records the users’ app usage by collecting data about browsing sessions. Each session is defined as a continuous time frame where the app is in the foreground on the smartphone. We used app session data to measure users’ engagement. To this purpose, conceptual decisions are required. One can use the number of sessions or the time spent on the app. The number of sessions itself can be per day or per week. The time spent on the app, in turn, can be computed per session or aggregated. We opted for time spent on the app aggregated over the week. \\
The original dataset provided 21283 app sessions of varying durations, up to 16 minutes. The app session data are presented in the second and third rows of Table \ref{tab:EdaWeeklyData}. For each user, we first aggregated each session duration on a daily scale, then on a weekly scale, to match the processing workflow previously described for scoring. We denoted such aggregated values as $d_i^{(k)}$ for all week indexes $i = 0, 1, ...$ and each user $k \in \{1, ..., 498\}$. Figure \ref{fig:EDA_engagem}, summarizing the behaviour of our users, shows that almost all policyholders look at the app mainly in the very first weeks of the program, and they spend few minutes weekly interacting with the app.

\subsection{Approaches for Data Analysis}\label{subsec:Approaches}

The first step of our analysis focused on improvement in driving behaviour. In the available literature, the notion of improvement has not yet been clearly defined. Weidner, Transchel and Weidner \cite[p.~214]{weidner2017telematic} claim that neither in science nor in practice there is a ``standardised method to achieve a clear `score' of driving behaviour''. Soleymanian, Weinberg and Zhu in \cite{soleymanian2019sensor} aggregate scores as mean values over weeks for the entire pool of policyholders and compare them among weeks, without distinctions among users and disregarding drop-off consequences. Yet, should an increase of the average score be considered an improvement, or are there more refined ways of defining it? Should we analyze the pool with average scores, or work at individual level? 

Unlike \cite{soleymanian2019sensor}, we decided to analyse improvement in driving behaviour for each individual policyholder and we introduced two different workflows. The first workflow, described in \ref{subsubsec:521}, explores for each user if there is an improvement of the initial driving style in any week of the period under consideration. The second one, discussed in \ref{subsubsec:522}, observes individual trends for the entire period.

In both cases, we split the users into four classes according to the quartile values of the 498 initial scores $s_0^{(k)}$. The \textit{merit-based} classes, labeled as `very-low', `medium-low', `medium-high' and `very-high', represent different initial scenarios for our analysis of coaching. In order to reliably measure improvement, we reasoned, it is implausible to include all drivers in an undifferentiated group, because the margins for improvement are of course very different for bad drivers with various critical issues, which can be addressed, than for drivers who already drive excellently, for whom there is little or no room for improvement. However accurate and effective it may be, coaching will have little effect on good drivers (those we include in the medium-high and very-high groups). At the same time, we can expect it to make a difference in the driving style of very-low and medium-low groups. Our analysis of improvement, therefore, differentially explores the improvement effects in the four groups of drivers with different skill levels. A precise description of merit-based classes can be found in Table \ref{tab:4classes_w0}. 

\begin{table}
 \caption{Statistical characterization of the four merit-based classes, used to divide the policyholders according to their initial score $s_0$.}
  \centering
  \begin{tabular}{lrrrr}
    \toprule
       & {very-low}     & {medium-low} & { medium-high}  & {very-high} \\
    \midrule
    Range for $s_0$ & [0, 0.4060]   & (0.4060, 0.6041]  & (0.6041, 0.7713]   & (0.7713, 1]   \\
    Median score         & 0.2915        &  0.4993           &  0.6853            & 0.8417   \\
    Number of users & 124           & 125               & 124                & 125      \\
    \bottomrule
  \end{tabular}
  \label{tab:4classes_w0}
\end{table}

In the following sections, we present the two approaches we used to investigate coaching effects based on two very different metrics to quantify the improvement of driving scores.

\subsubsection{Coaching effects over single weeks \label{subsubsec:521}}

In our first set-up we considered data points corresponding to the $s_i^{(k)}$ for i > 0. To study users’ behaviour after their initial week, we simply considered the difference:
\begin{equation}\label{eq:delta_ik}
    \delta_i^{(k)} = s_i^{(k)} - s_0^{(k)}, \qquad \forall i=1, 2, ...
\end{equation}
for each user $k$-th. In this case, we could independently study 3921 data points. As the scores are greater than zero, a positive value of $\delta_i^{(k)}$ denotes an enhancement in the driving style of the $k$-user in the $i$-th week with respect to his/her initial score. We consider only sufficiently high score increases given by $\delta_i^{(k)} > 0.05$  to be an \textit{improvement}, and we associate the corresponding data point to the \textit{deviation-based} class `Positive', relative to users with a positive coaching effect over single weeks. On the contrary, we cast the data points with $\delta_i^{(k)} < -0.05$ into the `Negative' class representing weekly driving sessions with behaviours worse than the initial one. Difference values $\delta_i^{(k)} \in [-0.05, +0.05] $ correspond to a null or very moderate variation of the driving score, and the corresponding data points are therefore associated to the `Null' class of driving sessions, with no relevant changes in the score. We note that the choice of the amplitude of the `Null'-related range is arbitrary, and we have reasonably set it as the 5\%-wide interval (as the scores are between 0 and 1, and $\delta$ can thus take values from -1 to +1).

From a methodological perspective, we remark that the definition of $\delta_i$ could also be based on the ratio $s_i^{(k)}/s_0^{(k)}$ instead of on the difference $s_i^{(k)} - s_0^{(k)}$. In that case, the three classes `Positive', `Negative', and `Null' would be defined around 1 by setting as stability range the thresholds based on the 5\% central interval.
In our analysis, we initially took both measures into account and then opted for the difference because, based on the dataset at our disposal, this approach allowed us to better discriminate the three deviation-based classes. We associated each $\delta_i^{(k)}$ to the cumulative duration $D_i^{(k)}$ of app usage during three weeks by summing the weekly duration (in seconds) as:
\begin{equation}\label{eq:Dik_eq2}
    D_i^{(k)} = d_{i-2}^{(k)} + d_{i-1}^{(k)} + d_i^{(k)}, \quad \forall i \ge 2, \forall k.    
\end{equation}
The reason for including the current week of analysis and the previous two weeks is that users’ behavioural patterns are not isolated within single weeks. Rather, they may be influenced by their interaction with the app in the previous weeks.


\subsubsection{Coaching effects over the entire period \label{subsubsec:522}}
\begin{figure}
    \centering
    \includegraphics[width=0.9\textwidth]{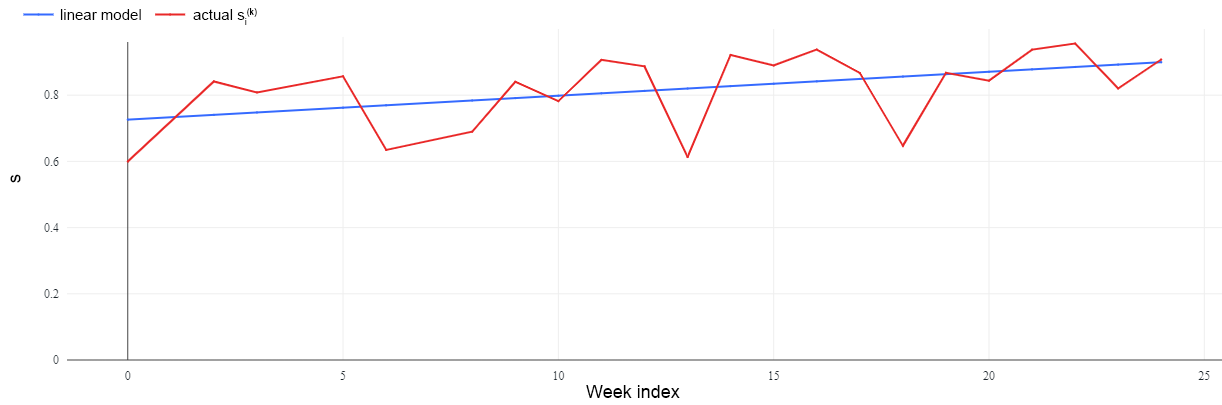 }
    \caption{Example of simple linear regression (blue line) computed on the weekly scores (in red) as function of the week index $i$, for a user with 25 tracked weeks.}
    \label{fig:LM_singleExample}
\end{figure}

The previous approach enabled us to take into account only temporary coaching effects, as the changes in each driver’s score do not correlate with the passing of time. Since we are interested in long-lasting improvements as well, we also analyzed the evolution of the score with respect to the week indices $i$-th for each user independently. We implemented linear regression models:
\begin{equation}\label{eq:LMformula}
    s(i) := \beta_0 + \beta_1 i
\end{equation}
explaining the score as the function $s$ of $i$, for each policyholder separately. An example of linear regression is reported in Figure \ref{fig:LM_singleExample}, for one (anonymous) user: on the horizontal axis, we read the index $i$ of the week since the user’s enrollment into the PHYD program, whereas the $[0, 1]$ vertical range is relative to the driving score function $s(i)$. His/Her positive slope coefficient $\beta_1$ denotes (on average) continuously improving driving performances over all tracked weeks.

Unfortunately, many users have few weekly scores, making the regression analysis not representative for them. We have thus discarded users with less than 8 scored weeks within the program and computed the regression coefficients for the remaining 212 users. 
We have performed the statistical analyses of the $\beta_1$ coefficients, relying on the popular p-values, for each driver. A low p-value 
suggests that the relationship between independent and dependent variables is statistically significant, i.e., the passing of time influences the driving score within the telematic program. Conversely, a high p-value indicates that this relationship could plausibly be due to random fluctuations in the data rather than an actual relationship between the variables. 
In this regard, it is well known that the conventional significance threshold of 0.05 may be unsuitable for studies with very small sample sizes, as in our case, and there is growing acceptance in various research fields to tolerate higher thresholds, such as 0.15 or 0.20 \cite{thiese2016pvalue}. Since our analysis focuses on capturing temporal trends rather than establishing a precise predictive model, we adopted 0.20. Thus, we interpret linear regressions with p-values $> 0.20$ as statistically not significant, and the corresponding users will be categorized as `Not Significant'.
On the contrary, for the users with p-value $\le 0.20$, we can capture patterns over time by looking at the value of their slope coefficient. These drivers are divided into three classes called `Negative', `Null', and `Positive', as in the previous approach. Specifically, drivers with $\beta_1 > 0.005$ are classified as `Positive', because their (sufficiently) positive slope denotes a long-term improvement in driving behaviour. Drivers with $\beta_1 < -0.005$ are classified as `Negative', as their scores decrease over the observed weeks, while users with  $-0.005 \le \beta_1 \le 0.005$ are classified as `Null' to denote that no practical relevant changes have been observed in their scores over time. In the following, we refer to these clusters of users as {\it slope-based} classes.

The variable quantifying the engagement of each $k$-th user to the app was computed as the mean of all the weekly durations $d_i^{(k)}$, i.e., as:
\begin{equation}\label{eq:Dk_eq4}
    D^{(k)} = {1 \over N^{(k)}} \sum_{i=0}^{N^{(k)}-1} d_i^{(k)},
\end{equation}
where $N^{(k)}$ is the user’s specific number of weeks within the telematics program. 
The choice to sum over all the weeks and divide by $N^{(k)}$ does not penalize policyholders who enrolled later.

\subsection{Limitations and remarks}\label{subsec:limitations}
Our research has some important limitations that must be taken into consideration, before moving on. 
One limitation is that we could not assess policyholders’ engagement in coaching programs that provide financial incentives. Whereas there is strong evidence that financial rewards are crucial for behavioural insurance programs \cite{stevenson2018effects, peer2020app, stevenson2021effect}, the database we explored refers to a policy without reward systems. 
The lack of financial rewards could explain the low digital engagement we observed in our database. If policyholders have no economic advantage, they have little or no incentive to make an effort to change their behaviour. Apparently, for people the motivation to save money is stronger than the motivation to change their behaviour for safety reason.

An additional limitation of our research lies in the impossibility of linking the data on engagement and improvement of driving style with the demographic characteristics of the users, which we could not access for privacy reasons. At the beginning of our research we wondered, for example, if there were differences in engagement between younger and older users (who presumably are less confident with digital technology), if the individual claim history was correlated with behavioural improvements, and other issues. Access to this data, of course, could provide both insurance scholars and companies with important insights.

There are two further remarks, both related to time.
As learned from \cite{soleymanian2019sensor}, few months may be sufficient to assess coaching effects on the insurance pool.
However, with a 9-month observation period, we could not ascertain whether and to what extent a `habituation' effect to the insurance program might occur, i.e., whether and to what extent time could affect engagement and, consequently, the effectiveness of the coaching program. Additionally, we could not examine the behaviour of policyholders over a period of time beyond the policy renewal threshold (one year and more), which could have been extremely informative, but exceeds the scope of this paper.

The limitations of our work, however, do not affect the value of the results, because the main objectives of our research are on the one hand to highlight the importance of engagement for the analysis and implementation of coaching programs, and on the other hand to propose a research methodology based on clear and unambiguous definitions of the concepts of engagement and coaching, which have so far been absent in the literature. The use of the available dataset serves us primarily to show how the proposed methodologies can be applied and how the results can be read/interpreted. The results we illustrate in session \ref{sec:results}, in fact, are not meant to be representative of the telematics motor insurance worldwide, but they offer clues to analyze and understand some real dynamics of that market.

%% file: aa_Results.tex
\section{Results and discussions}\label{sec:results}

We can now outline the main results achieved based on our dataset. 
We first focus on improvement merely. In a second step, we analyse engagement.

\subsection{Improvement}\label{subsec:Improvement}

In our opinion, a useful definition of improvement should consider two main issues: \textit{(i)} how to measure it and \textit{(ii)} what variation must be detected in order to speak of improvement and, consequently, of coaching. Focusing on the first issue, we point out that even slightly different measures (i.e., difference or ratio) lead to very different interpretations of single-week improvement.

\begin{figure}[ht]
    \centering
    \includegraphics[width=0.9\textwidth]{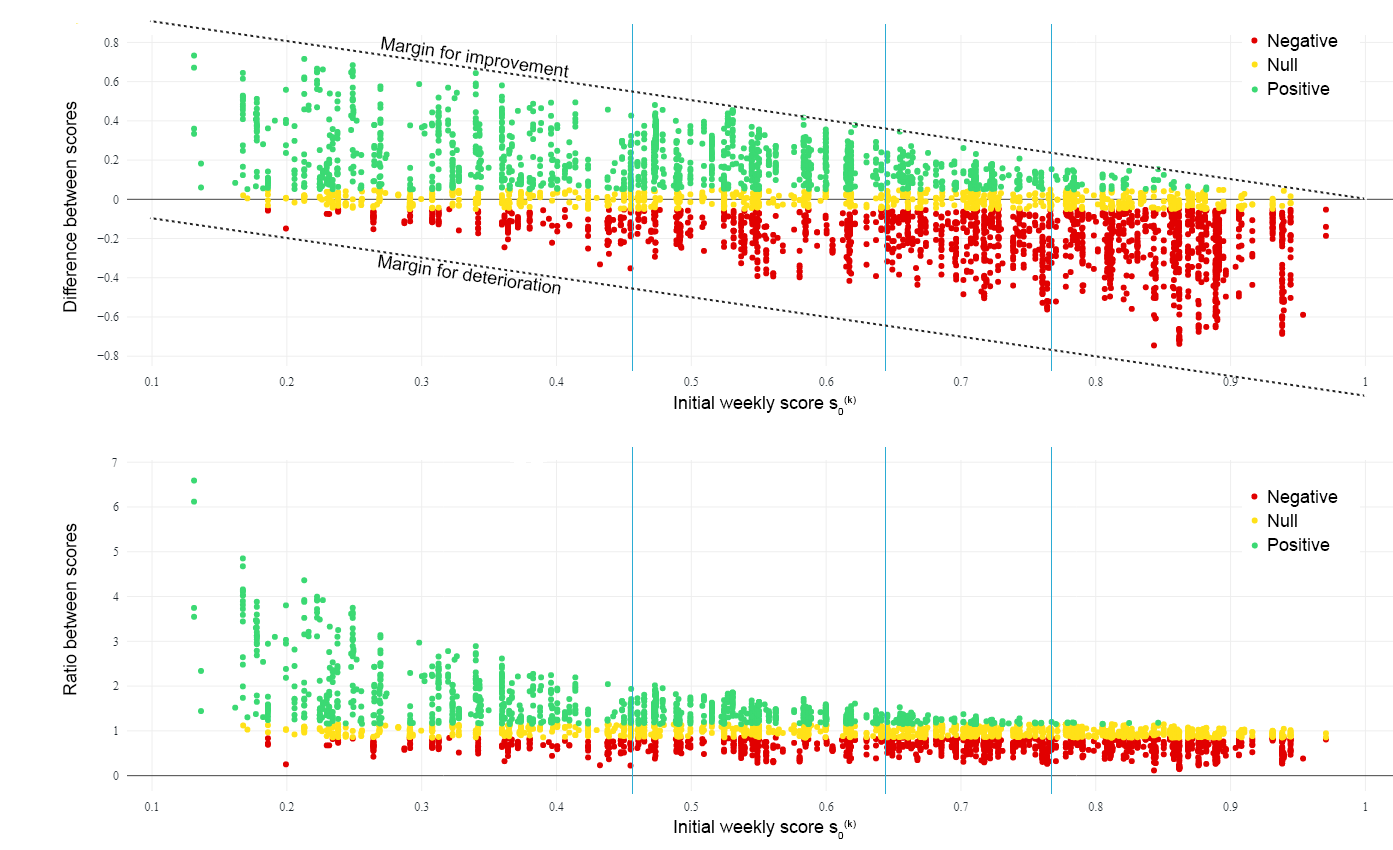}
    \caption{Scatterplots of data points given by $s_0^{(k)}$ initial scores, on the horizontal axis, and metrics for improvement quantification, on the vertical directions. The blue vertical lines denote the merit-based classes. On the top, the values of $\delta_i^{(k)}$ are given by difference as in Equation \ref{eq:delta_ik}; on the bottom, the values of $\delta_i^{(k)}$ are given by ratio.}
    \label{fig:Scatterplot_MarginImprov}
\end{figure}

In the scatterplots of Figure \ref{fig:Scatterplot_MarginImprov}, we look at our data points in terms of initial scores $s_0^{(k)}$ (on the horizontal axis) and $\delta_i^{(k)}$ values (on the vertical axis). In the first plot, the improvement values are computed as in Equation \ref{eq:delta_ik}; in the second plot, through the ratio between scores. In both plots, different colors highlight the three deviation-based classes. Especially in the first chart, one can see a wide fluctuation of scores independently from $s_0^{(k)}$, and the margins for deterioration or improvement seem to have been reached.

Table \ref{tab:DefImprovem} reports the cardinality of twelve data point clusters (derived by all possible combinations among the four merit-based and the three deviation-based classes) for both approaches. The two measures of single-week deviations from the initial score led to important differences: 35.50\% of data points (1390 out of 3921) denotes an improvement when it is measured as a difference, but the percentage decreases to 29.51\% (1158 out of 3921) if we adopt the ratio-based approach. In any case, our data indicates an effect of improvement in the single weeks for about one third of the cases.

\begin{table}[]
    \caption{Counting of data points in each merit-based and deviation-based class, according to two different methodologies for coaching analysis over single weeks.}
    \label{tab:DefImprovem}
    \centering
    \begin{tabular}{l rrr rrr}
\toprule
       &  \multicolumn{3}{c}{    Difference } & \multicolumn{3}{c}{Ratio } \\
   \cmidrule(lr){2-4} \cmidrule(lr){5-7}
       & Negative & Null & Positive   & \quad Negative & Null & Positive\\
    \midrule    
\multirow{2}{*}{very-low}      & 114 & 182 & 475       & 119 & 173 & 476 \\
  & (2.91\%) & (4.64\%) & (12.10\%) & (3.03\%) & (4.41\%) & (12.13\%) \\
   \arrayrulecolor{black!30}\midrule
\multirow{2}{*}{medium-low}    & 306 & 215 & 541       & 265 & 320 & 477 \\
  & (7.80\%) & (5.48\%) & (13.79\%) & (6.75\%) & (8.15\%) & (12.16\%) \\
   \arrayrulecolor{black!30}\midrule
\multirow{2}{*}{medium-high}   & 471 & 222 & 313       & 371 & 438 & 197 \\
  & (12.00\%) & (5.66\%) & (7.98\%) & (9.45\%) & (11.16\%) & (5.02\%) \\
   \arrayrulecolor{black!30}\midrule
\multirow{2}{*}{very-high}     & 759 & 262 & 64        & 527 & 550 & 8 \\
  & (19.34\%) & (6.68\%) & (1.63\%) & (13.43\%) & (14.02\%) & (0.20\%) \\
   \arrayrulecolor{black!30}\midrule
\multirow{2}{*}{\it Totals}        & \textit{1650}& \textit{881} & \textit{1390}     & \textit{1282} & \textit{1481} & \textit{1158} \\
  & \textit{(42.05\%)} & \textit{(22.45\%)} &\textit{ (35.50\%)} & \textit{(32.67\%)} & \textit{(37.74\%)} & \textit{(29.51\%)} \\
 \arrayrulecolor{black}\bottomrule
    \end{tabular}
\end{table}

If we understand improvement, instead, not as the result of single-week factors but as a positive change in driving behaviour over time, we look at the slope parameter $\beta_1$ of the regression model in Equation \ref{eq:LMformula} and at the slope-based classes. 
In Table \ref{tab:4classes_isSlope}, we report the cardinalities of the sixteen clusters defined according to the merit-based and the slope-based classes. First, we observe that only 39.62\% of our regressions are significant. Second, most of the policyholders have significant regression slopes that are negative or very close to zero, and only 15.15\% of the policyholders (30 out of 212) significantly enhanced their driving style. This means that if we observe the entire pool through individuals, on average no long-term coaching effect can be detected. However, most of the policyholders associated with the `very-low' class have improved their driving score over weeks (12 out of 16). Finally, we note that policyholders of the `medium-high' and `very-high' classes still tend to be underestimated, as in the previous analytic approach.

\begin{table}[h]
 \caption{Counting of data points in each merit-based and slope-based class, for coaching analysis over the entire period. }
  \centering
  \begin{tabular}{l rrrrr}
    \toprule
       & {very-low}     & {medium-low} & {medium-high}  & {very-high} & \textit{Totals}  \\
    \midrule
\multirow{2}{*}{Not Significant}        & 28    & 35    & 30    & 35   & {\it 128}  \\
                & (13.21\%) & (16.51\%) & (14.15\%)   & (16.51\%) & {\it (60.38\%)}\\     
                 \arrayrulecolor{black!30}\midrule
\multirow{2}{*}{Negative}              & 3    & 10    & 15    & 14   & {\it 42}  \\
                & (1.42\%) & (4.72\%) & (7.08\%)   & (6.60\%) & {\it (19.81\%)}\\     
                 \arrayrulecolor{black!30}\midrule
\multirow{2}{*}{Null}          & 1    & 4    & 4    & 3   & {\it  12}   \\
                & (0.47\%) & (1.89\%) & (1.891\%)    & (1.42\%) & {\it (5.66\%)}\\    
                 \arrayrulecolor{black!30}\midrule
\multirow{2}{*}{Positive}        & 12    & 8    & 7    & 3    & {\it 30}  \\
                & (5.66\%) & (3.77\%) & (3.30\%)   & (1.42\%) & {\it (15.15\%)}\\ 
                 \arrayrulecolor{black!30}\midrule
\multirow{2}{*}{\it Totals}    & {\it 44} & {\it 57} & {\it 56} & {\it 55}  & {\it 212} \\
                & \textit{(20.75\%)} & \textit{(26.89\%)} & \textit{(26.42\%)}  & \textit{(25.94\%)} & {\it (100\%)}\\
  \arrayrulecolor{black} \bottomrule
  \end{tabular}
  \label{tab:4classes_isSlope}
\end{table}

\begin{figure}
    \centering
    \includegraphics[width = 0.45\textwidth]{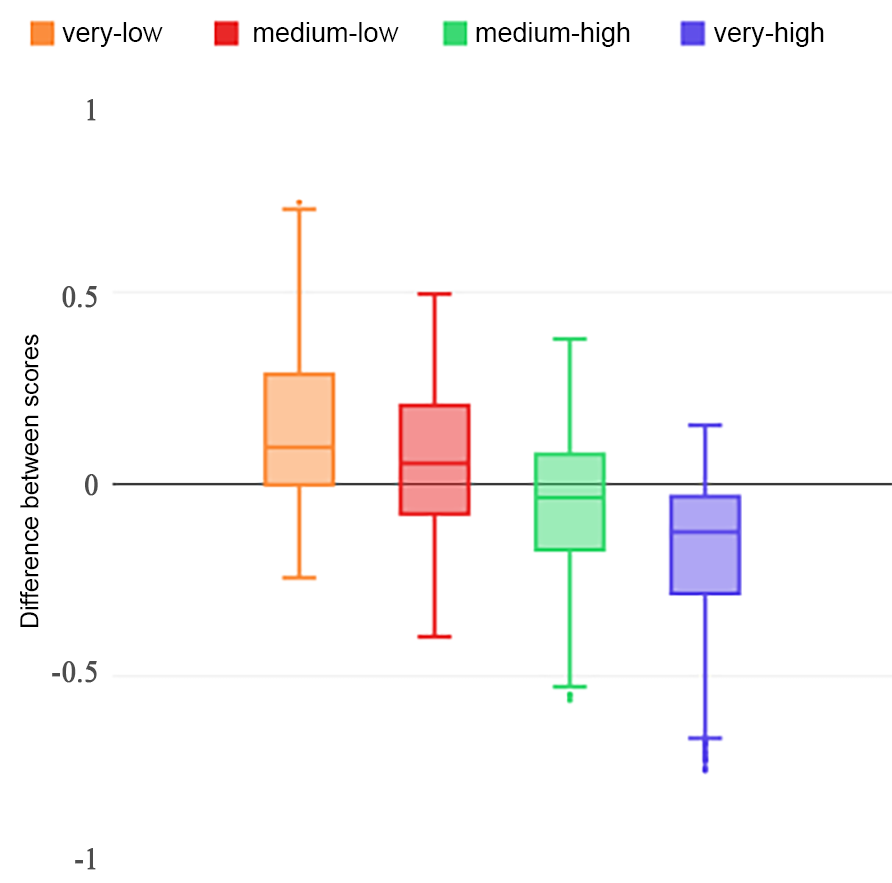}
    \caption{Boxplots of the $\delta_i^{(k)}$ values of Equation \ref{eq:delta_ik} }
    \label{fig:Boxplot}
\end{figure}

The division of policyholders into the four merit-based classes introduced in \ref{subsec:Approaches}  allows us to describe our heterogeneous pool more accurately, taking into account the initial level of driving ability. We can then better interpret the real complexity and variety of the results. In Figure \ref{fig:Boxplot}, we separately report through boxplots the $\delta_i^{(k)}$ values (computed with differences, as in Equation \ref{eq:delta_ik}) for the merit-based classes.\\
For `very-low' and `medium-low' classes, the positive median values (represented with horizontal lines inside the orange and red boxes) show an improvement in the driving style of policyholders. Moreover, for the `very-low' class, the entire orange box lies over the zero line, denoting that in 75\% of cases there was an increase in the score. On the other hand, as expected, the purple box is below the zero line, coherently with the numerous occurrences of red points on the right-hand side of the first scatterplot in Figure \ref{fig:Scatterplot_MarginImprov}. Due to their limited margins for improvement, good drivers are prone to worse scores. This suggests that the `very-high' class should not be penalised for slight score deterioration and, more generally, that any improvement or deterioration of the score should be assessed relative to the starting conditions and not in absolute terms.

\subsection{Engagement}\label{subsec:Engag_62}

\begin{figure}
    \centering
    \includegraphics[width=0.7\textwidth]{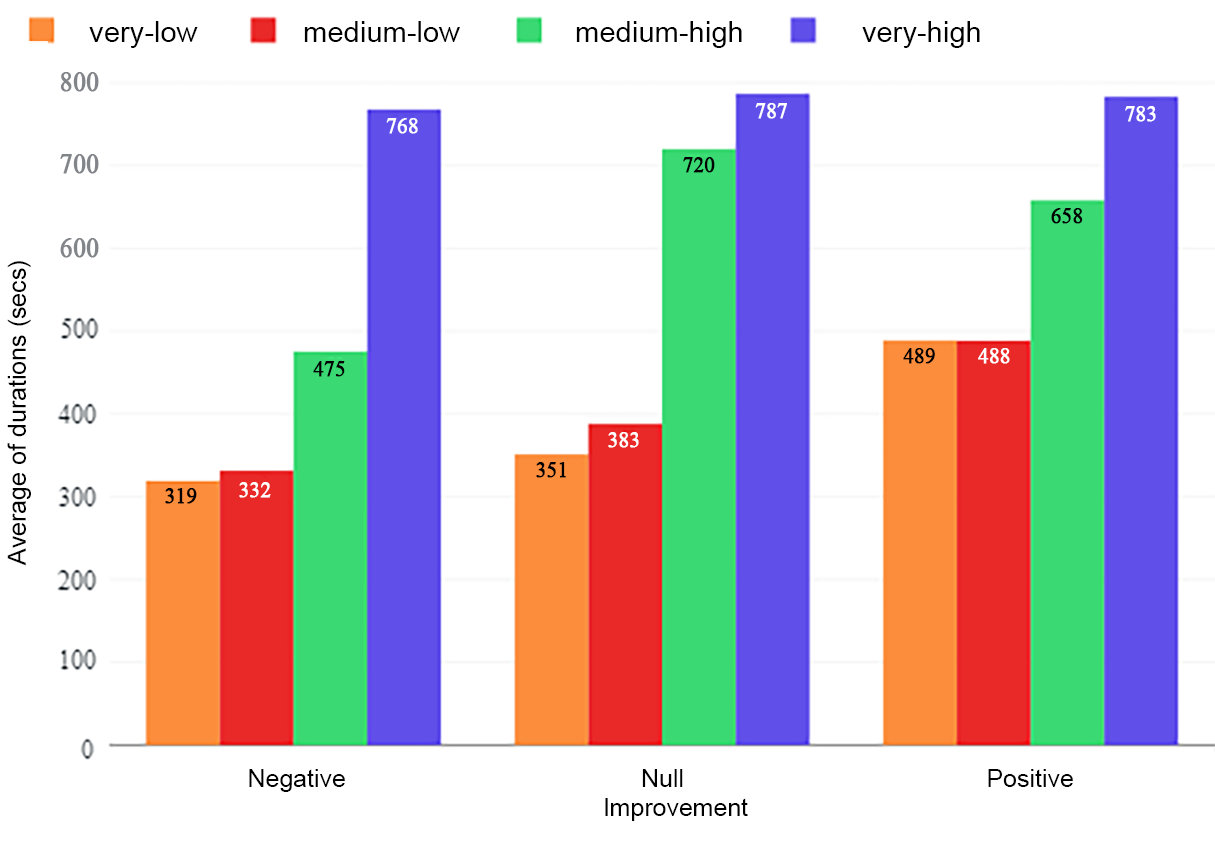}
    \caption{Average of the engagement metric $D_i^{(k)}$ defined in Equation \ref{eq:Dik_eq2}, for the merit-based and the improvement-based classes in single weeks.}
    \label{fig:BarplotEngag_SingleWeek}
\end{figure}

\begin{figure}
    \centering
    \includegraphics[width=0.9\textwidth]{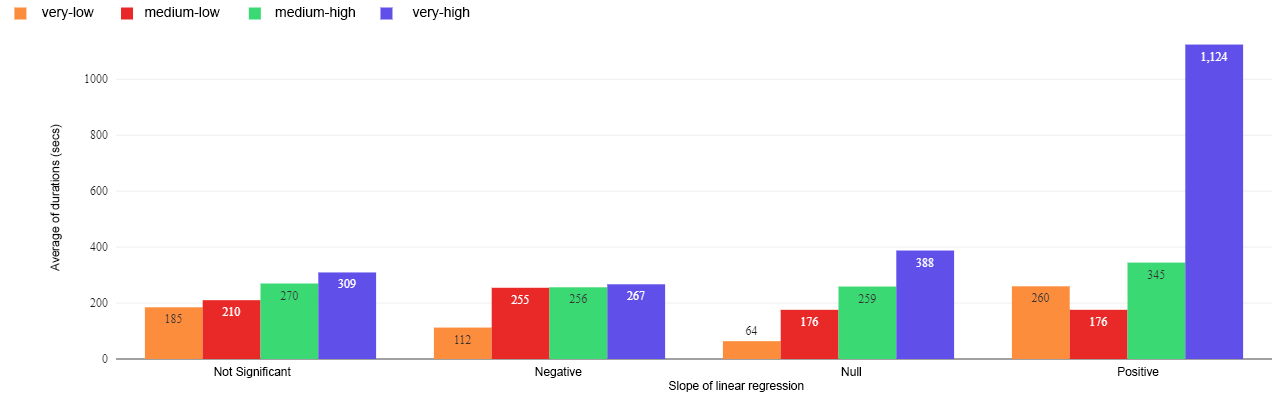}
    \caption{Average of the engagement metric $D^{(k)}$ defined in Equation \ref{eq:Dk_eq4}, for the merit-based and the improvement-based classes over the entire period.}
    \label{fig:BarplotEngag_LM}
\end{figure}

In the last step of our analysis, we face the conceptual challenges concerning the idea of engagement. As pointed out in the Introduction, the definition of engagement is problematic. 
Since by ‘engagement’ we mean the active dealing of policyholders with the information fed back by the insurance company, we measured the duration of each user’s interaction with the telematics app summing the seconds spent on the app according to the definitions given in Equations \ref{eq:Dik_eq2} and \ref{eq:Dk_eq4}, which refer respectively to single weeks and the entire period. The results of the two assessments for all the considered clusters are reported in Figure \ref{fig:BarplotEngag_SingleWeek} and Figure \ref{fig:BarplotEngag_LM}.


In both cases, it emerges that the users of the `very-high' class (blue bars) are the most engaged with the telematics app, independently from their cluster. In most deviation- and slope-based classes, the subdivision into the four merit-based clusters reveals that the higher the initial scores, the longer the time spent on the telematics app. This holds also for the users with no statistical relevant regressions.
In addition, we observe a correlation between engagement and score variation. In Figure \ref{fig:BarplotEngag_SingleWeek}, especially in the `very-low' and `medium-low' classes (orange and red bars), the customers manifesting positive coaching effects in single weeks are the most engaged ones, whereas less engaged customers do not improve their driving scores. In  Figure \ref{fig:BarplotEngag_LM}, this trend is only partially confirmed. However, the (three) drivers in the `very-high' class that improve their scores show very strong engagement over the entire period, well above the other groups.

Overall, these results show that there is an interplay between engagement and improvement, but this connection is not strongly evident in all cases. We suppose that the weak interplay is due to the poor engagement characterizing our dataset. Figure \ref{fig:EDA_engagem} shows that the number of users looking at the app is very low if compared to the number of users enrolled in the telematics program, and the interest in checking information feedback and scores rapidly decreases over time. Moreover, users do not have long sessions with the app, which can suggest that they do not properly read all informative pages.

%% file: aa_Concl.tex
\section{Conclusions}\label{sec:Concl}

The investigation presented in this work contributes to clarify the definition of engagement and the related methodological challenges that must be addressed when exploring engagement in proactive insurance policies. 
Behavioural telematics data promise to change both the traditional insurance business model and the interaction between insurers and policyholders, providing a solution to the classical problem of moral hazard \cite[p.~1231]{van2006making}. 
The paradox underlying moral hazard is that policyholders are less incentivised to take precautionary measures because they are insured \cite{arrow1978uncertainty, stiglitz1983risk}. In proactive insurance, the argument goes, if individual behaviour could be observed and either rewarded or penalised depending on exposure to dangers, this could affect the propensity of policyholders to control their behaviour and the problem of moral hazard could be, if not removed, at least mitigated. 
In PHYD insurance policies, telematics of course does not control moral hazard by directly steering individual behaviour. The only control that might take place is a kind of self-control, that can be based on the motivation to improve behaviour, but also on the awareness of being tracked, or on the mere possibility of earning financial incentives \cite{cevolini2022coaching}. To this purpose engagement plays a crucial role: our research shows that improvement effects are higher in policyholders who actively interact with the app.

Our findings, however, also show that engagement cannot be taken for granted. Many policyholders do not look at the app at all, and the ones who do it tend to have short and superficial sessions. The usage of any app is time-consuming and can be annoying over time. This finding is confirmed by research in behavioural data-based health insurance, that is, in the so-called pay-as-you-live (PAYL) insurance policies \cite{lupton2016diverse, ruckenstein2017beyond, gorm2019episodic, tanninen2022uncertain}.
If the goal of PHYD insurance policies is to improve driving style, policyholders must first be motivated to engage with the app in order to be motivated to change their behaviour. Is there anything insurance companies can do to increase the level of engagement? 

Insurance companies could expand the functionality of the app and make the interaction with it more appealing. For example, effective engagement can depend on the app usability, but also on short messages and notifications proposing challenges to achieve in order to improve driving behaviour, earn points, and be rewarded. If these messages and notifications were personalised, coaching could be custom-made and the app-based interaction with the insurance company could be more exciting.  
Such interaction is itself a kind of behaviour producing second-order behavioural data that can be recorded and used strategically. Insurance companies, therefore, could also implement second-order coaching strategies with the goal of improving engagement, besides the first-order strategies aiming at improving driving behaviour.

The open question is how insurance companies can make use of these second-order scores. In principle, the level of engagement could be taken into account in the scoring process as well. Including engagement into the score, however, is not without risks. One could reward the most engaged users with points they earn when they interact with the telematics app. But users who knew that their score also depends on this interaction could be inclined to game the system, learning to improve their score rather than learning to improve their driving behaviour.
Indeed, information fed back to the drivers can affect how people behave, but also how people deal with their behaviour. 

For example, a father who is late in driving his daughter to school might turn off his mobile phone and turn it back on when he drives calmly from school to work. The interaction with the app becomes strategic, if not even opportunistic. Sure, a single event does not affect a long exposure, and by means of crossing data the lack of the trip can be detected. On the other hand, a systematic misuse can be associated to fraud. Regardless of these two extreme cases, it would be fine to reward users who stay tuned, but a reward based on the level of engagement with the app would work only if it were not disclosed. However, can insurance companies not disclose how they calculate the score?

These difficulties can be interpreted as flaws of behavioural insurance policies, but also as confirmation of the crucial role of engagement, which could give rise to innovative approaches. In mundane everyday life, instead of optimizing their driving style, users are rather prone to optimize their interaction with the tracking device. Instead of keeping their driving behaviour under control, users keep their tracking behaviour under control. This could explain why long-term engagement is so difficult to achieve and could be the starting point for insurance companies to implement more complex UBI business models.